\documentclass[12pt]{article}
\usepackage{natbib}
\usepackage{graphics}
\usepackage{amsmath,amsfonts,amssymb,bm,dsfont,amsbsy}
\usepackage[english]{babel}
\usepackage[displaymath]{lineno}
\usepackage{color}
\usepackage{latexsym}
\usepackage{amsthm}
\usepackage{multirow}
\usepackage{multicol}
\usepackage{float}
\usepackage{wrapfig}
\usepackage[latin1]{inputenc}
\usepackage{times}
\usepackage{graphicx}
\usepackage{geometry}
\theoremstyle{definition}

\begin{document}

\begin{center}

\Large {\bf Measuring heterogeneity in urban expansion via spatial entropy}\par

\bigskip

\normalsize{Linda Altieri, Daniela Cocchi, Giulia Roli}\\ \small{Department of Statistical Sciences, University of Bologna, via Belle Arti, 41, 40126, Bologna, Italy} \par

\bigskip
\end{center}

\begin{abstract}
The lack of efficiency in urban diffusion is a debated issue, important for biologists, urban specialists, planners and statisticians, both in developed and new developing countries. Many approaches have been considered to measure urban sprawl, i.e. chaotic urban expansion; such idea of chaos is here linked to the concept of entropy. Entropy, firstly introduced in information theory, rapidly became a standard tool in ecology, biology and geography to measure the degree of heterogeneity among observations; in these contexts, entropy measures should include spatial information. The aim of this paper is to employ a rigorous spatial entropy based approach to measure urban sprawl associated to the diffusion of metropolitan cities. In order to assess the performance of the considered measures, a comparative study is run over alternative urban scenarios; afterwards, measures are used to quantify the degree of disorder in the urban expansion of three cities in Europe. Results are easily interpretable and can be used both as an absolute measure of urban sprawl and for comparison over space and time.\end{abstract}
\begin{quotation}
\bigskip
\end{quotation}

\section{Introduction}
\label{sec:intro}

Urban sprawl is characterized by uncontrolled development of cities into surrounding areas, which has aroused wide social focus because its induced urbanization is inefficient, dispersed and may impede sustainable development. Rapid urban growth is quite alarming worldwide, and the importance of
conducting research on this topic is strongly felt \citep{johnson, ewing08, review}. Although an accurate definition of urban sprawl is still debated, the general consensus is that urban sprawl is characterized by `unplanned and uneven pattern of growth, driven by a multitude of processes and leading to a inefficient resource utilization' \citep{bhatta}. More definitions appear in  \cite{jaeger10} and focus on the negative consequences of sprawl. The negative impacts of urban sprawl concern many aspects, not only for human life quality (e.g. increased costs and time for transportation), but also for the environment. The dispersion of urban areas increases pollution, waste of soil and soil consumption. This endangers ecosystems and species, and reduces the availability of land for agriculture, water bodies, forests and other natural areas \citep{eea_foen}. In addition, urban sprawl does not foster climate changes mitigation, even if variations in climate do not immediately fit with the velocity of uncontrolled urbanization. Any spatial planning strategy has a different impact on climate change \citep{bart, stone12}, but the standard consequences of uncontrolled urbanization concern strong precipitation events, additional heat due to increased emission of carbon dioxide and, in particular,  heat island effects.

In Europe, urban sprawl is an increasing issue \citep{eea_sprawl, couch, eea_foen}, which can be evaluated according to several viewpoints. For instance, \cite{eea_foen} stress that the spatial configuration of the built up areas is a fundamental component of urban proliferation. Different arguments in EEA reports point out the impact of urban sprawl: the negative effects mentioned before are even more evident if the costs for future generations are taken into account, and are related to the ideas of fragmentation, degradation and consequences on ecosystems.

The literature about sprawl is voluminous (e.g. \citeauthor{torrens}, \citeyear{torrens}; \citeauthor{bhatta}, \citeyear{bhatta}; \citeauthor{ent_urban}, \citeyear{ent_urban}; \citeauthor{ewing}, \citeyear{ewing}; \citeauthor{oues}, \citeyear{oues}); the quantification of the phenomenon develops according to different routes that keep into account alternative formulations of demographic, social and economic variables. This is partly due to the difficulties of a unique definition. Moreover, characterisation of sprawl in the literature is often narrative
and subjective, and measurement largely depends on data, to the point that existing studies yield contrary results for the same cities in several cases  \citep{torrens, bhatta}. The basic sprawl indicator is the low level of population density over an area; in other words, it declares whether an unnecessary waste of urbanised land occurs. Alternatively, sprawl may be defined in cost terms, as in \cite{benfield}, or by ratios of urban growth \citep{ewing}. A lot of sprawl measures are indeed based on ratios: relative measures quantify attributes of
urban growth and can be compared among cities, among different zones within a city, or across
different times \citep{bhatta}. Such ratios are easy to interpret and receive a lot of discussion, but are statistically poor. In order to capture different aspects that are related to sprawl, \cite{jiang} proposed an integrated urban sprawl measure that combines 13 indices; unfortunately, the final measure requires extensive inputs of temporal data, and does not mention any threshold to characterise a city as sprawling or non-sprawling.

Among the proposals for urban sprawl measurements, there is a number of spatial or landscape metrics, that have long been used in landscape ecology. Landscape metrics aim at evaluating the spatial pattern of land cover classes or entire landscape mosaics of a geographic area. Indeed, the urbanization of a territory can be assessed according to the exhibited pattern of land cover classes: a  sprawled city is in contrast with a compact one, with 'empty' (i.e. non-urban) spaces and scattered urban areas denoting inefficient development. Consequently, land cover and land use data are particularly suitable for urban sprawl measurements. Such data usually are vector (polygonal) or raster (pixel) spatial data coming from remote sensing images, where the territory is classified in a finite number of categories according to the prevailing land use, after a definition about what land use classes are considered as urban or non-urban. Then, the pattern of urban areas and its evolution over time can be exploited to quantify urban dispersion as lack of spatial clustering (compactness) of the urban patch. For an approach to sprawl measurement based on a comparative use of Moran's I with land use data, see \cite{nostro-ecolin}.

Two aspects need to be considered when assessing the presence of urban sprawl with land use data. First, the objective is to detect lack of compactness, i.e. heterogeneity, in the territory by looking at the spatial dispersion of the urban tissue. Secondly, the variable of interest, land use, is qualitative and unordered; this aspect is particularly critical for environmental statistics, as it limits the set of tools for data analysis. The need to deal with categorical variables and the detection of heterogeneity in the territory support the idea of employing entropy measures. Shannon's entropy is used in several fields, such as geography, ecology, biology, to assess the heterogeneity of a population over an area. Ecological concepts, such as evenness and richness, are strictly related to heterogeneity, and entropy represents the utmost index to measure heterogeneity in a dataset. In the context of urban sprawl, entropy has proved to be a stringent measurement tool \citep{yeh}, and is still a widely used technique, suitable for integration of remote sensing and GIS \citep{bhatta, cora, remotesens}. While entropy succeeds in working with qualitative variables and quantifying the heterogeneity of a dataset, it suffers from the drawback of not considering the role of space as a source of heterogeneity in determining the variable outcomes. Indeed, Shannon's entropy is computed based on the proportions of the land use classes, not on their spatial configurations, and two territories with the same proportions and very different degrees of compactness for the urban tissue have the same entropy value; the same holds for territories with different area size. Shannon's entropy is not affected by size, shape and number of sub-areas of a spatial territory, while a spatial metric for urban sprawl should be. The urban sprawl issue is tightly bond to the spatial location of land use data. Therefore, appropriate studies of sprawl which make use of entropy measures should introduce spatial information.

Over the past decades, two main approaches have been adopted to include spatial information into an entropy measure. Extending Theil's work (\citeyear{theil}), \cite{batty74, batty76,  batty10} introduced the first approach by defining a spatial entropy measure accounting for unequal space partition into sub-areas. In \citeyear{karlstrom}, this proposal was modified by \citeauthor{karlstrom} to satisfy the property of additivity, i.e. decomposing of the global index into local components. The main drawback of this approach is that such entropy can only be computed for a binary variable. Moreover, the local terms are not entropies and do not possess the properties of the global one,  and results are heavily affected by the selected area partition. Nevertheless, the approach proves to be informative in the context of urban sprawl.  
The second approach to spatial entropy is based on a suitable transformation of the study variable that accounts for the distance between realizations (co-occurrences). The main proposals have been made by \cite{oneill}, \cite{contagion}, \cite{leibovici09} and \cite{leibovici14}, but all these distance-based  measures do not enjoy the additivity property and rely on the choice of a single distance without capturing the behaviour of the studied variable. A recent work by \cite{nostro} fulfils desirable properties by proposing a set of spatial entropy measures starting from the co-occurrence approach and focusing on pairs of realizations. The resulting entropy is decomposed into the information due to space and the remaining information brought by the variable itself once space is considered. The proposal preserves additivity and disaggregates results, allowing for partial and global syntheses.

The properties of spatial entropy measures make them an appealing tool to evaluate urban sprawl from a spatial perspective. A spatial entropy measure is sensitive to the spatial dispersion of urban patches over an area and may be able to separate the heterogeneity of land use data due to the lack of spatial compactness from the heterogeneity due to other components. They enjoy basic desirable properties of any spatial index \citep{anselin}, i.e. the additivity between local and global results. They also receive interpretation and are suitable for delivering results across different areas of expertise.

The main aim of this work is to adopt entropy based tools for measuring urban sprawl in terms of spatial compactness or dispersion. If sprawl is considered as a negative condition, and is measured  by means of spatial entropy, a low level of entropy is desirable, i.e. a non-chaotic (compact) urban configuration. We present a thorough assessment of the advantages and disadvantages of a selection of spatial entropy measures which have not been employed in the context of urban sprawl measurement yet, both with a comparative study on simulated data and via a 
case study on three European cities. The simulation study compares spatial entropy values across representative urban configurations: the monocentric, the polycentric and the decentralized city. In addition, the resulting ranges of entropy values may be used as reference intervals for comparison to real case studies, as in the application, where we propose an example of comparison over space and time, that can be extended as wished.
Our results can be combined with measures integrating relevant demographic, social or economical variables affecting urban sprawl.

The motivating case study comes from official European land use data. We selected two time points, 1990 and 2012, for the commuting belts of three cities in Europe: Bologna (also studied in \citeauthor{nostro-ecolin}, \citeyear{nostro-ecolin}), Eindhoven and Lublin. They belong to countries with different levels of urban sprawl \citep{eea_sprawl}. 

Though spatial entropy is applied to the specific issue of urban sprawl, the techniques illustrated in the present paper may be used for any phenomenon whose spatial distribution and heterogeneity is of interest. Their evaluation is relevant for climate and meteorology studies, e.g. the spatial distribution of metereological phenomena, for ecological purposes, e.g. species distribution (\citeauthor{nostro}, \citeyear{nostro}), for general landscape and geographical studies, for the assessment of environmental risks, e.g. earthquakes and wildfires, for atmospheric studies, e.g. polluting substances, for disease mapping. 

In the present paper, in Section \ref{sec:reviewspace} we revisit the works by \cite{batty74} and \cite{karlstrom} under a unified statistical framework. We also illustrate the approach of \cite{nostro} with a special focus on its use in urban sprawl studies. In Section \ref{sec:sim}, we build a simulation study, which compares, evaluates and discusses the performance of the two approaches for spatial entropy measures under different urban scenarios. This is useful both for further applications, since the study covers the main urban configurations, and as a contribution to the statistical theory of spatial entropy measures. In Section \ref{sec:applic}, the measures are applied to the case study; this constitutes a further practical contribution to the discussion on urban sprawl. Some concluding remarks can be found in Section \ref{sec:disc}.

This work is implemented in \texttt{R} \citep{R}. It makes use of the packages \texttt{sp} \citep{sp}, \texttt{spatstat} \citep{spatstat} and dependencies, and of the recent package \texttt{SpatEntropy} \citep{spatentropy} now available on CRAN.

\section{The use of spatial information in entropy measures}
\label{sec:reviewspace}

In many environmental and urban studies, the definition of entropy measures coincides with Shannon's formula: given a categorical variable $X$ with $I$ possible outcomes, the entropy is
\begin{equation}
H(X)=\sum_{i=1}^{I}p(x_i)\log\left(\frac{1}{p(x_i)}\right)
\label{eq:shann}
\end{equation}
where $p(x_i)$ is the probability of the $i$th outcome and $\log\left(1/p(x_i)\right)$ is the information function, which measures the information brought by outcome $x_i$ \citep{coverthomas}. Entropy is a non-negative quantity, which measures the average 'information' or 'surprise' concerning an outcome of $X$. The more the categories of $X$ are equally likely, the higher the entropy; if a category of $X$ is far more likely than others, the entropy is low, as one can predict the behaviour of $X$ and data do not carry much information. Thus, entropy synthesizes the heterogeneity of $X$ outcomes in a single number; data with very different spatial configurations but the same probability mass function for $X$ share the same entropy. In the context of urban sprawl, this is not desirable. For example, an area which is partly urbanized and partly rural may be compact, with an urban nucleous and rural surroundings, or dispersed, i.e. sprawled, with many small scattered urban areas. Shannon's entropy does not detect the difference in the two patterns and returns the same value if the proportion of urbanized and non-urbanized territory is the same across the two configurations.

For this reason, an extension to spatial entropy is needed. The seminal attempt to extend (\ref{eq:shann}) into a spatial entropy measure developed by \cite{batty74} is presented in Section \ref{sec:batty}; its most relevant extension, proposed by \cite{karlstrom} is sketched in Section \ref{sec:kc}. A recent approach to spatial entropy, proposed by \cite{nostro}, is in Section \ref{sec:nostra}. 
All measures assume a peculiar meaning in the analysis of urban sprawl. They are very suitable in distinguishing the desirable situation of urban compactness from urban sprawl.

Most spatial entropy measures make use of the concepts of spatial adjacency and neighbourhood. The notion of neighbourhood is linked to the assumption that occurrences at certain locations are influenced, in a positive or negative sense, by what happens at surrounding locations, i.e. their neighbours. 
The system can be represented by a graph \citep{bondy}, where each location is a vertex and neighbouring locations are connected by edges. The simplest way of representing a neighbourhood system is via an adjacency matrix: for $G$ spatial units, $A=\{a_{gg'}\}_{g,g'=1,\dots,G}$ is a square $G\times G$ matrix such that $a_{gg'}=1$ when there is an edge from vertex $g$ to vertex $g'$, and $a_{gg'}=0$ otherwise; in other words, $a_{gg'}=1$ if $g' \in \mathcal{N}(g)$, the neighbourhood of area $g$. Its diagonal elements are all zero by default. In this work, spatial units may be pixels or polygons, defined via representative coordinate pairs, such as the area centroids, which are used to measure distances and define what units are neighbours. In the remainder of the paper, the word 'adjacent' is used accordingly to mean 'neighbouring', i.e. connected in the graph, while the word 'contiguous' is used for pixels or polygons sharing a border on the map, i.e. a topological contact.

\subsection{Towards additive spatial entropy}
\label{sec:spatent}

\subsubsection{Batty's spatial entropy}
\label{sec:batty}

A very appreciable attempt to include spatial information into Shannon's entropy starts from a reformulation of (\ref{eq:shann}). The categorical variable $X$ is recoded into $I$ dummy variables, each identifying the occurrence of a specific category of $X$, where, by construction, $p_i=p(x_i)$.

This approach is proposed by Batty (\citeyear{batty74, batty76}) to define a spatial entropy which extends Theil's work (\citeyear{theil}). In a spatial context, a phenomenon of interest $F$ occurs over an observation window of size $T$ partitioned into $G$ areas of size $T_g$. This defines $G$ dummy variables identifying the occurrence of $F$ over a generic area $g$, $g=1, \dots, G$.
Given that $F$ occurs somewhere over the window, its occurrence in area $g$ takes place with probability $p_g$, where $\sum_g p_g=1$. The intensity is obtained as $\lambda_g=p_g/T_g$, where $T_g$ is the area size, and is assumed constant within each area. Shannon's entropy of $F$ may be written as
\begin{equation}
H(F)=\sum_{g=1}^G p_g \log \left(\frac{1}{p_g}\right)=\sum_{g=1}^G \lambda_gT_g \log \left(\frac{1}{\lambda_g}\right)+\sum_{g=1}^G \lambda_gT_g \log \left(\frac{1}{T_g}\right).
\end{equation}
\cite{batty76} shows that the first term on the right hand side of the formula converges to the continuous version of Shannon's entropy \citep{renyi}, namely the differential entropy, as the area size $T_g$ tends to zero. The differential entropy is rewritten in terms of $p_g$, giving Batty's spatial entropy
\begin{equation}
H_B(F)=\sum_{g=1}^G p_g \log \left(\frac{T_g}{p_g}\right).
\label{eq:spaten}
\end{equation}
It expresses the average amount of information brought by the occurrence of $F$ over the areas, and includes $T_g$ that accounts for unequal space partition. Analogously to Shannon's entropy, which is high when the $I$ categories of $X$ are equally represented over a (non spatial) data collection, Batty's entropy is high when the phenomenon of interest $F$ is equally intense over the $G$ areas partitioning the observation window (i.e. when $\lambda_g=\lambda$ for all $g$). Batty's entropy $H_B(F)$ reaches a minimum value equal to $\log(T_{g^*})$ when $p_{g^*}=1$ and $p_g=0$ for all $g\ne g^*$, with $g^*$ denoting the area with the smallest size. The maximum value of Batty's entropy is $\log(T)$, reached when the intensity of $F$ is the same over all areas, i.e. $\lambda_g=1/T$ for all $g$.
This maximum value does not depend on the area partition, nor on the discrete or continuous nature of $F$, but only on the size of the observation window.
When $T_g=1$ for each $g$, $H_B(F)$ is a Shannon's entropy of $F$ equivalent to (\ref{eq:shann}), and the index ranges accordingly in $[0,\log(G)]$.

When the target is to measure urban sprawl, $F$ denotes the presence of urbanization. A high level for Batty's entropy is not desirable, as it indicates constant urban intensity, i.e. scattering of urban patches across regions, denoting sprawl. A low level, on the contrary, indicates that some areas in the window have a very high urban density (usually, the city centre) while others tend not to present urbanization (i.e. the outside areas). Therefore, when Batty's entropy is low the city is compact and a scarce level of sprawl is present, which is interpreted as a positive condition. 

\subsubsection{A LISA version of Batty's spatial entropy}
\label{sec:kc}

A challenging attempt to introduce additive properties and to include the idea of neighbourhood in Batty's entropy index (\ref{eq:spaten}) is due to \citeauthor{karlstrom} (\citeyear{karlstrom}), following the LISA theory \citep{anselin}.
\citeauthor{karlstrom}'s entropy index $H_{KC}(F)$ starts by weighting the probability of occurrence of $F$ in a given spatial unit $g$, $p_g$, with its neighbouring values:
\begin{equation}
\widetilde{p}_g=\sum_{g'=1}^G a_{gg'}p_{g'}.
\label{eq:karl_ptilde}
\end{equation}
Then, an information function is defined, fixing $T_g=1$, as $I(\widetilde{p}_g)=\log\left(1/\widetilde{p}_g\right)$.
In this proposal, the elements on the diagonal of the adjacency matrix $A$ are non-zero, i.e. each area neighbours itself and enters the computation of $I(\widetilde{p}_g)$. \citeauthor{karlstrom}'s entropy index is
\begin{equation}
H_{KC}(F)=\sum_{g=1}^G p_g\log\left(\frac{1}{\widetilde{p}_g}\right).
\label{eq:karl}
\end{equation}
The maximum of $H_{KC}(F)$ does not depend on the choice of the neighbourhood and is $\log(G)$. As the neighbourhood reduces, i.e. as $A$ tends to the identity matrix, $H_{KC}(F)$ coincides with Batty's spatial entropy (\ref{eq:spaten}) in the case of $T_g=1$ for all $g$.
The sum of local measures $p_gI(\widetilde{p}_g)$ constitutes the global index (\ref{eq:karl}), preserving the LISA property of additivity.

One major disadvantage of (\ref{eq:spaten}) and (\ref{eq:karl}) is that a categorical variable $X$ with $I>2$ outcomes cannot be used, since only one category enters the measure. In other words, $F$ may be a specific category of $X$, say $F=X_i^*$, and $H_{KC}(X_i^*)$ is computed to assess the spatial configuration of the realizations of $X_i^*$. Thus, for a categorical $X$, $I$ different $H_{KC}(X_i^*)$ are computed, but no way is proposed to synthesize them into a single spatial entropy measure for $X$. Moreover, the local components are not entropy measures themselves. Lastly, conclusions are affected by the choice of the area partition. Nevertheless, Batty's and \citeauthor{karlstrom}'s approach is expected to be helpful in the context of urban sprawl, and is assessed in Sections \ref{sec:sim} and \ref{sec:applic}.

\subsection{Spatial entropy based on a transformation of the study variable}
\label{sec:nostra}

A second way to build a spatial entropy measure consists in defining a new categorical variable $Z$, where each realization identifies pairs $\{x_i,x_j\}$ of occurrences of $X$ over space \citep{oneill, contagion, leibovici09}. Such change of variable is crucial in a spatial context, since space is now considered via the distances between observations forming a pair. 
For $I$ categories of $X$, the new variable $Z$ has $R=(I^2+I)/2$ categories. The attention moves from the computation of (\ref{eq:shann}), namely $H(X)$, to an index of the same form, Shannon's entropy of $Z$, $H(Z)$. 

\cite{nostro} follow the approach based on $Z$ and introduce a second discrete variable $W$, that represents space by classifying the distances at which the two occurrences take place. These classes $w_k$, with $k=1,\dots,K$, cover all possible distances within the observation window. The definition of the classes is exogenous and depends on the study at hand (\citeauthor{nostro}, \citeyear{nostro}). Each distance category $w_k$ implies the choice of a corresponding adjacency matrix $A_k$, which identifies pairs where the two realizations of $X$ lie at a distance belonging to the range $w_k$. 

Thanks to the introduction of $W$, the entropy of $Z$ may be decomposed as
\begin{equation}
H(Z)=MI(Z,W)+H(Z)_{W}
\label{eq:shannZ_add}
\end{equation}
following the fundamentals of Information Theory \citep{coverthomas}: the first term $MI(Z,W)$ is known as mutual information and measures the amount of the entropy of $Z$ which is explained by its relatioship with $W$, while the second term $H(Z)_{W}$ is the conditional, or residual, entropy, quantifying the remaining amount of entropy of $Z$ once the effect of $W$ is removed. In a spatial context, the two terms acquire a new meaning: $MI(Z,W)$ is the quantity of interest in this context, and is called spatial mutual information, because $Z$ identifies pairs of categories of spatial observations and $W$ collects categories of distances where pairs can take place. Spatial mutual information quantifies the part of entropy of $Z$ due to the spatial configuration $W$; for the same reason, $H(Z)_{W}$ is the spatial global residual entropy, quantifying the information brought by $Z$ after space has been taken into account.
The more $Z$ depends on $W$, i.e. the more the realizations of $X$ are spatially associated, the higher the spatial mutual information. Conversely, when the spatial association among the realizations of $X$ is weak, the entropy of $Z$ is mainly due to spatial global residual entropy. 

When it comes to sprawl, the variable of interest $X$ has categories urban/non-urban, and $Z$ identifies pairs with the three possible unordered combinations of urban/non-urban areas (urban/urban, urban/non-urban, non-urban/non-urban). A compact city represents the situation where the $X$ outcomes should be highly positively correlated. In such case, spatial mutual information tends to be high, because urban areas generally have urban neighbours, while non-urban areas have non-urban neighbours; space plays a relevant role in determining the entropy of $Z$.
The overall value of $MI(Z,W)$, however, is negatively influenced by what happens at large distance ranges, where usually scarce correlation is present. Hence, spatial mutual information for the whole dataset may approach zero even when a compact pattern occurs.

The variable $W$ helps in overcoming this drawback, since the two terms forming $H(Z)$ can be further decomposed. Indeed, $K$ subsets of realizations of $Z$ are available,
denoted by $Z|w_k$; for all the distance classes $w_k$ a set of $K$ conditional distributions is obtained, that sum up to the two components of (\ref{eq:shannZ_add}). When measuring urban sprawl, this means that the degree of compactness of a city may be quantified at different distance ranges, which can help in understanding the extent and seriousness of the sprawl phenomenon.

From Information Theory, spatial mutual information:
\begin{equation}
MI(Z,W)=\sum_{k=1}^{K}p(w_k)PI(Z|w_k)=\sum_{k=1}^{K}p(w_k)\sum_{r=1}^{R}p(z_r|w_k)\log{\left(\frac{p(z_r|w_k)}{p(z_r)}\right)}
\label{eq:PI_add}
\end{equation}
is a weighted sum of partial terms $PI(Z|w_k)$, each quantifying the contribution of the $k$th distance range to the spatial mutual information between $Z$ and $W$. In other words, each partial term measures the degree of association (compactness) in the city pattern at each distance range. The focus is expected to be on short distance ranges, where the difference between a compact city and a dispersed one is more evident. By exploring these terms, an indication of the degree of sprawl can be provided.

Analogously, 
\begin{equation}
H(Z)_W=\sum_{k=1}^{K}p(w_k)H(Z|w_k)=\sum_{k=1}^{K}p(w_k)\sum_{r=1}^{R}p(z_r|w_k)\log{\left(\frac{1}{p(z_r|w_k)}\right)},
\label{eq:residZWadd}
\end{equation}
where the partial residual entropy terms measure the partial contributions to the entropy of $Z$ due to sources other than the spatial configuration. As regards sprawl, a great value for $H(Z|w_k)$, especially at short distance ranges, is a hint for urban dispersion.

The additive terms in  (\ref{eq:PI_add}) and (\ref{eq:residZWadd}), together with their sums, constitute a rich set of spatial entropy measures. In particular, spatial mutual information has theoretical support to be considered a reliable method for measuring urban heterogeneity. It is able to maintain the information about the categories of $X$ by exploiting the trasformed variable $Z$, to consider different distance ranges simultaneously, to quantify the overall role of space, and to be easily interpretable. A comparative study for different urban configurations is developed in what follows, in order to verify its ability to detect sprawl.

\section{Spatial entropy measures on simulated urban settings}
\label{sec:sim}

\begin{figure}
\centering
\caption{Examples of the three urban scenarios: monocentric, polycentric, decentralized.}\label{fig:1}
\includegraphics[width=.9\textwidth]{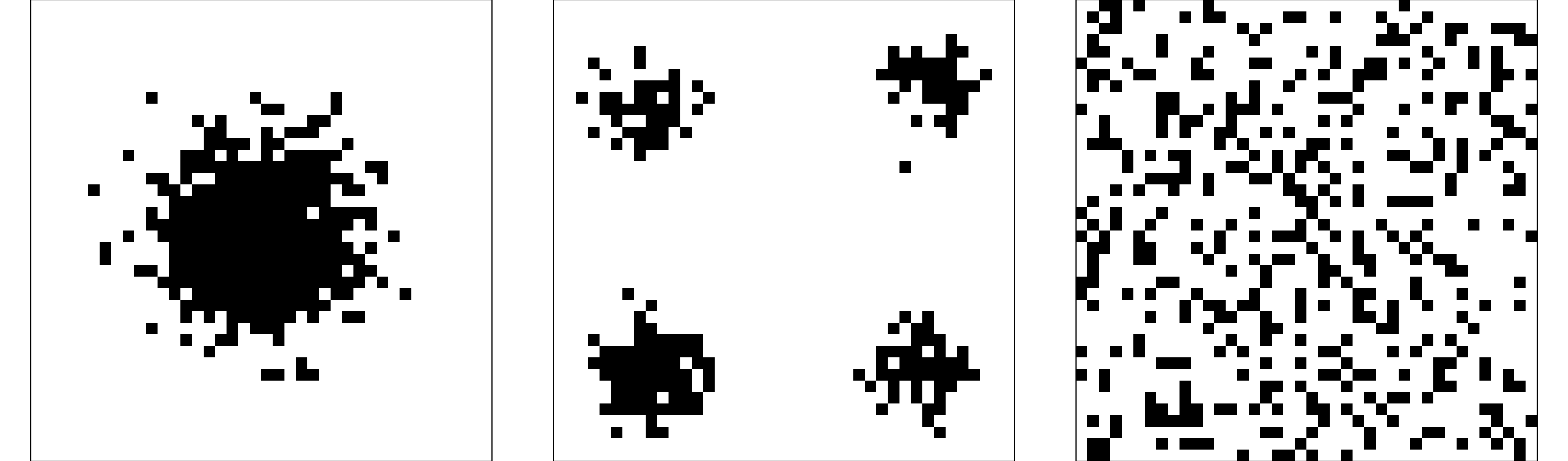}
\end{figure}

The flexibility and informativity of the spatial entropy indices discussed in Section \ref{sec:reviewspace} are assessed with a comparative study, which aims at understanting the differences between the two approaches over three main urban configurations. Following \cite{tsai}, they are identified as monocentric city, polycentric city and decentralized city. The monocentric city is considered the most positive situation as regards the urban pattern; the polycentric city is an intermediate, less compact, situation which may suffer from sprawl; the decentralized configuration is concerned by the sprawl issue. An example of the three settings is shown in Figure \ref{fig:1}. 
\begin{center}
Insert Figure \ref{fig:1} about here
\end{center}
The three scenarios initially come as point patterns on a square area of size 100. The monocentric and polycentric scenarios are generated from the intensity function of a Thomas process \citep{spatstat}, i.e. a Poisson cluster point process,  with one cluster for the monocentric case and four clusters for the polycentric case. The decentralized pattern is generated following the intensity function of a homogeneous Poisson process. For the three urban scenarios, 1000 datasets are simulated. Then, the point patterns are gridded and turned into raster data: each data matrix is 40$\times$40 pixels, so that each pixel has side 0.25 and area size 0.0625. The binary variable is $X$ with $x_1=$ urban and $x_0=$ non-urban. Consequently, $Z$ has 3 categories: $z_1=\{$urban, urban$\}$, $z_2=\{$urban, non-urban$\}$, $z_3=\{$non-urban, non-urban$\}$. Parameters for data generation are such that, for each of the 1000 realizations, the number of urban and non-urban pixels is the same across the three scenarios. This way, Shannon's entropy would not be able to distinguish among the configurations, while we check how the measures of Section \ref{sec:reviewspace} succeed in detecting sprawl.

\subsection{Batty's and \citeauthor{karlstrom}'s entropy}
\label{sec:simres_batty}

\begin{figure}
\centering
 \caption{Two options for area partition in Batty's entropy over an example of monocentric dataset. Left panel: 20 random areas; right panel: 5 concentric rings.}    \includegraphics[width=.6\textwidth]{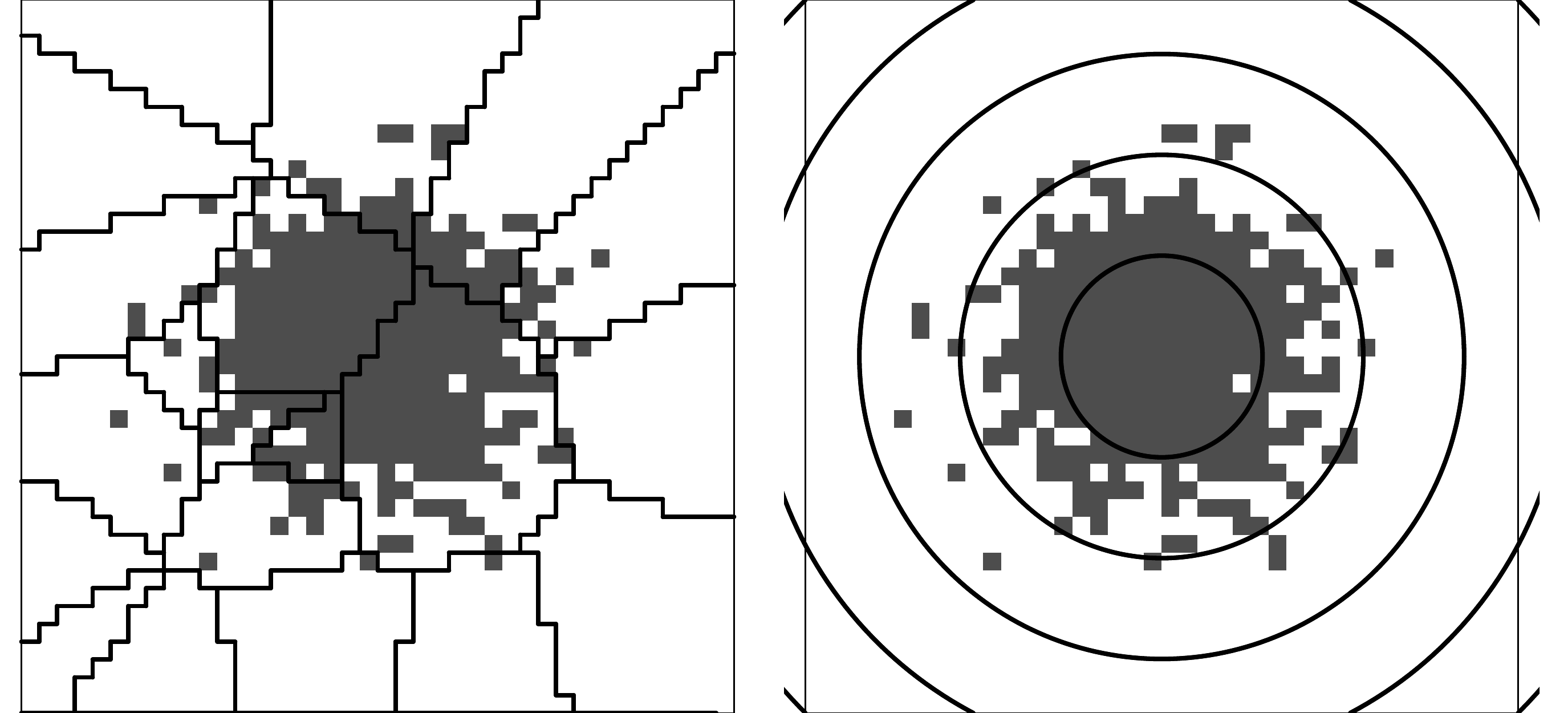}
         \label{fig:batty_part}
\end{figure}

\begin{figure}
\centering
 \caption{Results for Batty's entropy over the three urban scenarios, 1000 simulations, with the two partition options: 20 random areas (left panel), 5 concentric rings (right panel).}    \includegraphics[width=.6\textwidth]{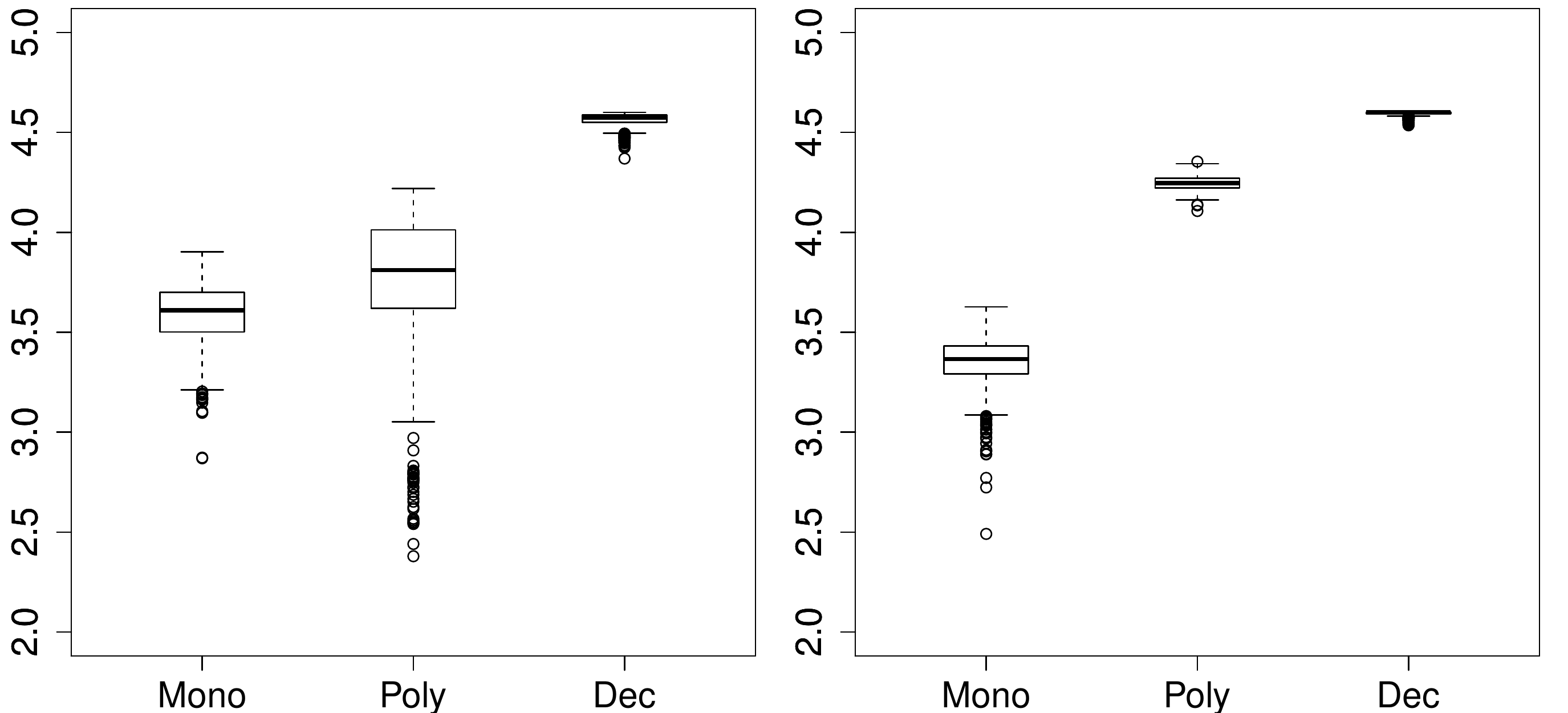}
         \label{fig:batty_2opt}
\end{figure}

Entropies of Section \ref{sec:spatent} cannot be computed directly on the pixel grid, since only one realization of $X$ occurs over each pixel, while such entropy measures need a population of pixels over a wider area. The phenomenon $F$ is here defined as the occurrence of urban pixels, i.e. $F=X_1$. Since these measures are substantially affected by the area partition, we check two different options for splitting the observation area into sub-areas. Firstly, the observation area is partitioned into $G1=20$ areas of different size, by randomly generating 20 centroids over the area and then performing a Dirichlet tessellation, i.e. assigning each pixel to the area with the closest centroid. A second option, more appropriate in the context of urban sprawl, is to partition the observation area into concentric sub-areas, which can give a better idea of city expansion into surrounding areas. We choose $G2=5$ annuli, defined by concentric rings, with the same width, i.e. the same difference between the radius of the outer ring and the one of the inner ring. The annuli center is the observation area centroid, and their width is chosen so that they cover the whole area. The two options are shown in Figure \ref{fig:batty_part} for a monocentric dataset. For both options, the probabilities $p_g$ are estimated in each of the $1000$ simulations as the proportions of urban pixels over the sub-areas.
\begin{center}
Insert Figure \ref{fig:batty_part} about here\\
Insert Figure \ref{fig:batty_2opt} about here
\end{center}
Batty's entropy for the three scenarios and the two partition options is shown in the boxplots Figure \ref{fig:batty_2opt}. The measure is able to distinguish among the three urban configurations as regards spatial entropy: the monocentric, non sprawled case has a lower entropy distribution, the polycentric scenario returns intermediate values and the decentralized pattern returns a distribution of very high entropy values, close to Batty's maximum. The distinction between the decentralized scenario and the other two is evident with both partition options, but the concentric one, more suitable in an urban context, shows that the ranges for all three scenarios do not overlap: this case can be used as a reference set in real studies. For comparison purposes, relative values (i.e. divided by the maximum $\log(100)$) should be used: the lowest value for the decentralized pattern is 0.985, thus considered a benchmark for urban sprawl.

For \citeauthor{karlstrom}'s entropy, different possibilities for the neighbourhood distances between the sub-areas' centroids are considered, in order to quantify $I(\widetilde{p}_g)$. For partition option 1, three neighbourhoods are set using the 5th percentile, first quartile and median of the distribution of distances among the $G1=20$ areas' centroids; they are equal to 
$nd_{11}=1.473$, $nd_{12}=3.654$ and $nd_{13}=5.335$. For option 2, four neighbourhoods are possible over the 5 annuli, i.e. up to the $j$th farthest area, $j=1, \dots, 4$. We name them $nd_{21}=1\mbox{Ann}$, $nd_{22}=2\mbox{Ann}$, $nd_{23}=3\mbox{Ann}$ and $nd_{24}=4\mbox{Ann}$, where $jAnn$ means 'up to the $j$th farthest annulus'. The estimates of $\widetilde{p}_g$ are computed for the 3 neighbourhoods of the first case and the 4 neighbourhoods of the second case as averages of the neighbouring estimated probabilities. 

\begin{figure}
\centering
 \caption{Results for \citeauthor{karlstrom}'s entropy over the three urban scenarios, 1000 simulations, with the two partition options: 20 random areas (higher panels), 5 concentric rings (lower panels), at different neighbourhood distances.}    \includegraphics[width=\textwidth]{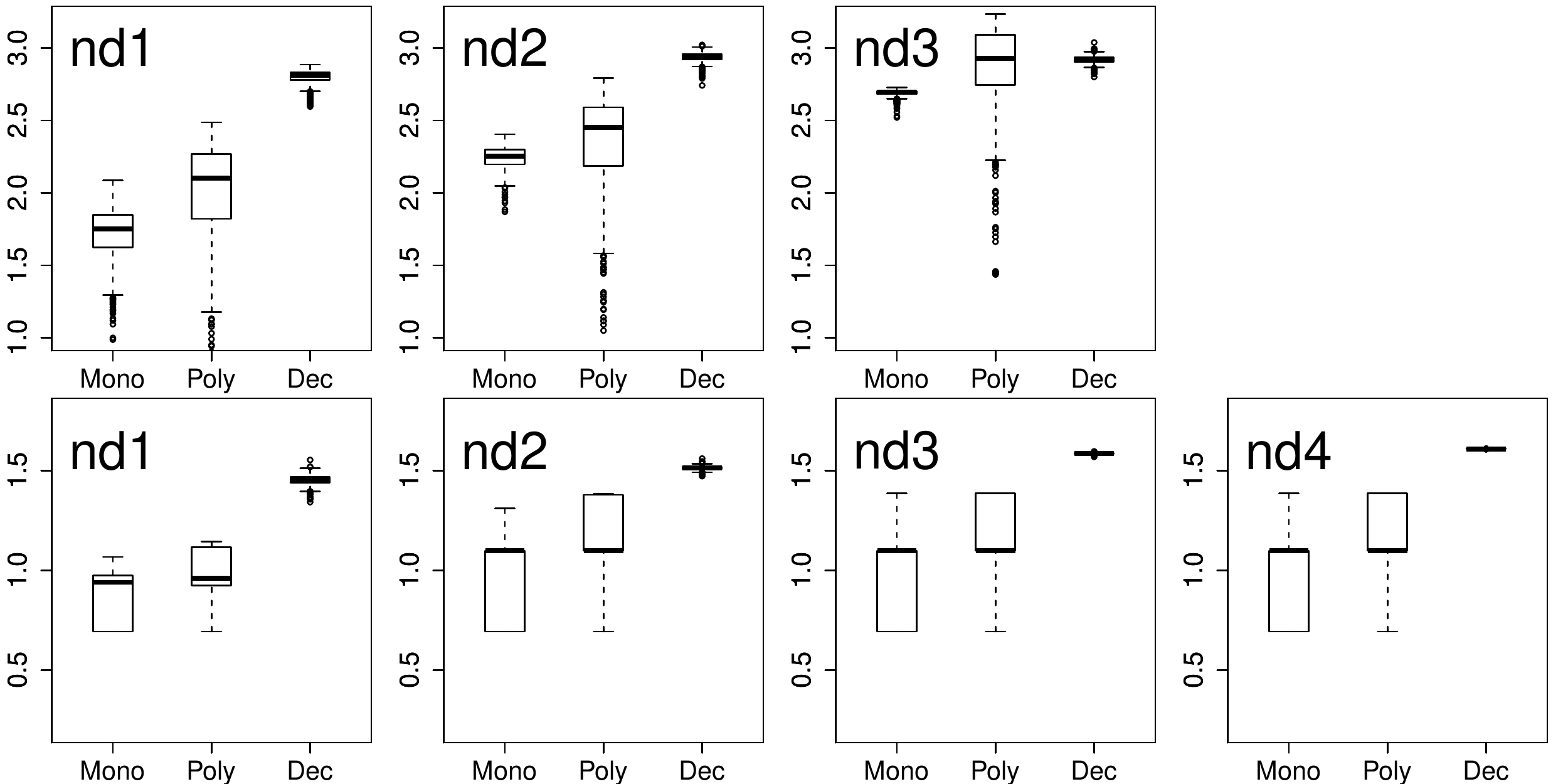}
         \label{fig:karl_2opt}
\end{figure}

Results for \citeauthor{karlstrom}'s entropy are shown in Figure \ref{fig:karl_2opt}, again for the three urban configurations and all neighbourhood options. This entropy measure distinguishes the first two urban patterns from the decentralized one when the neighbourhood distance is small. The second partition option (lower panels) yields again more suitable results. However, the interquartile ranges tend to overlap, therefore, the measure is not generally able to determine what type of urban configuration is present. While \citeauthor{karlstrom}'s extension to Batty's entropy is interesting from a theoretical point of view because of the LISA-type properties, it does not seem to provide major advantages in practical situations. Widening the neighbourhood (from left to right panels in both lines of Figure \ref{fig:karl_2opt}) tends to increase all entropy values and to generate confounding among patterns. It should also be remembered that the results shown in the panels represent choices that are separately, not jointly, computed, with the consequence of obtaining limited information in applied case studies.
\begin{center}
Insert Figure \ref{fig:karl_2opt} about here
\end{center}

The overall limit of this approach is that results are heavily affected by the choice of the area partition.

\subsection{Spatial mutual information and residual entropy}
\label{sec:simres_Z}

For the computation of the entropy set of Section \ref{sec:nostra}, breaks for the distance ranges must be chosen, where the distance concerns pairs of pixels, not sub-areas as in Section \ref{sec:kc}, and is measured between pixel centroids. [1.13] Two options are considered in the simulation study. The first one is motivated by the tradition of spatial statistics, where the so called 4 nearest neighbour system (i.e. pixels sharing a border) and the analogous 12 nearest neighbours system are of standard use \citep{anselin}. Accordingly, the first two distance breaks chosen for option 1 are $w_{11}=[0,0.25]$ and $w_{12}=]0.25, 0.5]$, where 0.25 is the distance between contiguous pixels' centroids; the remaining breaks are $w_{13}=]0.5, 1.25]$, i.e. up to 5 pixels along the cardinal directions, and $w_{14}=]1.25, d_{max}]$, $d_{max}=13.789$ being the maximum distance between pixels within the observation area. This way, the first three classes are quite small, while the last one is very large. In the measurement of urban sprawl, the focus is on what happens at small distance ranges, where a lack of spatial association, i.e. a high presence of pairs of type $\{$urban, non-urban$\}$, indicates dispersion, thus sprawl. Therefore, detailed results are needed for small distances, while aggregate results are enough at large distances. The second option follows the same criterion as the neighbourhood distance choice in Section \ref{sec:simres_batty} [1.13]: the empirical distribution of pixel distances is computed, and the breaks are chosen as the 5th, 25th and 50th percentile: $w_{21}=[0,1.346]$, $w_{22}=]1.346,3.260]$, $w_{23}=]3.260,5.130]$, $w_{24}=]5.130,13.789]$. The global values are not affected by the choice of the $w_k$ and can be further modified if wished. Pairs are built for each distance range $w_k$ according to the specific adjacency matrix $A_k$, which identifies the pairs of pixels at a distance that belongs to the $k$th range. The rule of moving rightward and downward is adopted along the observation window in order to identify neighbouring pairs, to avoid double counting. Then, each $p_{Z|w_k}$ is estimated using proportions for the three categories of $Z$ at the specific distance range.

\begin{figure}
\centering
 \caption{Spatial partial information for the three urban scenarios, 1000 simulations. First option for the distance ranges in the higher panels, second option in the lower panels.}
\includegraphics[width=\textwidth]{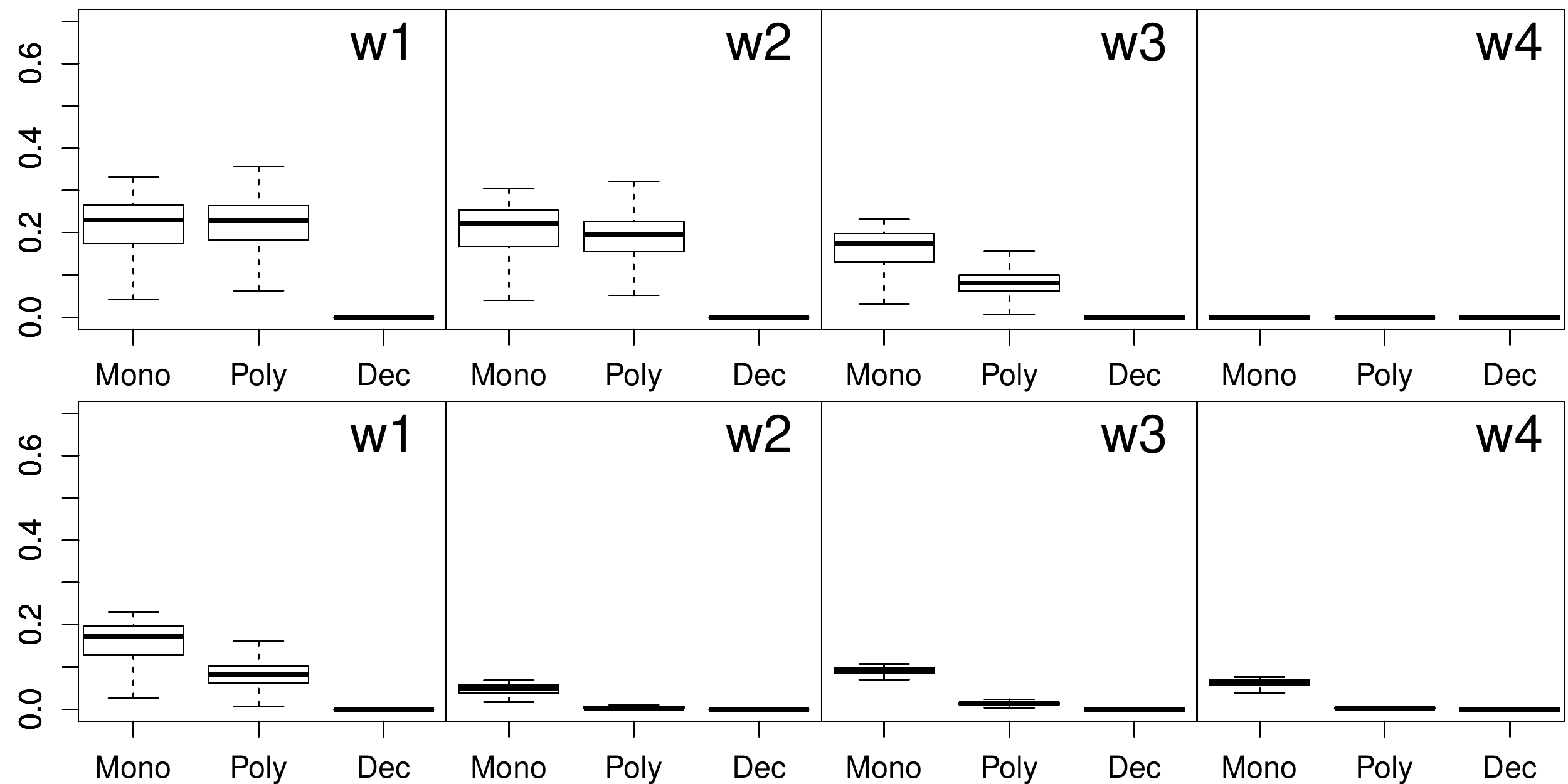}
         \label{fig:partmut_X2}
\end{figure}

Shannon's entropy computed for $X$ or $Z$ is the same, and does not depend on the spatial configuration. Thus, entropy $H(Z)$ can be safely used to evaluate the entropy of the variable of interest, i.e. urbanization, with the additional advantage of considering distances between urban/non-urban pixels. Spatial mutual information illustrates how the role of space is detected following the three considered spatial configurations. Since the main focus of this work is on the contribution of the partial terms, rather than on the global value, spatial partial information terms are shown in Figure \ref{fig:partmut_X2} for the two distance class options. 
\begin{center}
Insert Figure \ref{fig:partmut_X2} about here
\end{center}
For the first distance option (higher panels) an appreciable influence of space is detected at very short distances for the first two spatial patterns (mono- and polycentric), while the difference between the two becomes more evident as distance increases. For spatial mutual information, we ought to obtain the same results for mono- and polycentric cities at $w_{11}$: when only contiguous pixels are considered, the spatial behaviour of the two configurations is the same. The second option (lower panels) has wider distance classes: class $w_{21}$ aggregates former classes $w_{11}$, $w_{12}$ and $w_{13}$. Here, the distinction among configurations is very evident for $w_{21}$. For further distance ranges, the role of space is only detected in the monocentric scenario. No mutual information is detected at any distance over the decentralized patterns, where no spatial structure is present and space does not help in explaining the data behaviour. Spatial mutual information can be interpreted as a sprawl detector: a high mutual information value implies positive association among urban areas and positive association among non-urban ones, and indicates a compact urban expansion. Another appreciable advantage of this measure is that information at different distance ranges is available and knowledge is gained about the data spatial behaviour. The boxplots in Figure \ref{fig:partmut_X2} can be used as reference intervals for assessing real case studies, since no overlap occurs between a compact and a sprawled situation. At very broad distance classes (right hand side panels) the lack of distinction among patterns is expected and is of scarce interest in sprawl studies. The choice of the classes does not affect the global result, unlike the choice of Batty's area partition. 

Results are not shown for spatial residual entropy, as its interpretation is symmetrical to the interpretation of spatial mutual information: a high proportion of residual entropy at short distance ranges denotes urban sprawl. We believe spatial mutual information to be the key component of entropy for drawing conclusions on sprawl. Beyond enjoying the theoretical properties summarized in Section \ref{sec:reviewspace}, spatial mutual information proves to be effective in measuring urban sprawl and distinguishing among scenarios.

\section{Measuring urban sprawl in Europe via spatial entropy}
\label{sec:applic}

\begin{figure}
\centering
 \caption{From left to right, Eindhoven, Lublin and Bologna together with their commuting belts, in 1990 (higher panels) and 2012 (lower panels). }    \includegraphics[width=\textwidth]{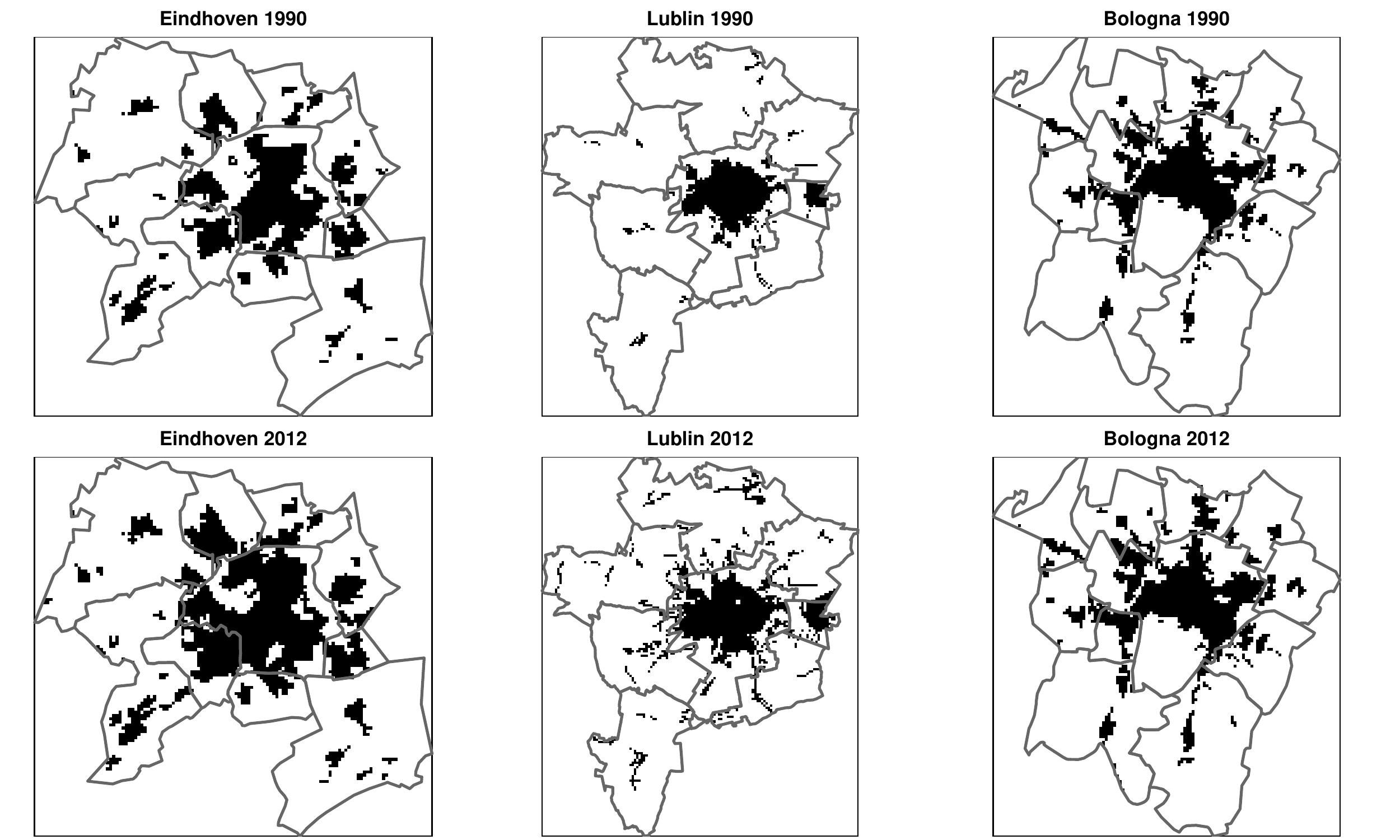}
         \label{fig:bolo}
\end{figure}

The case study comes from official European sources. Land use data for the entire European territory are made available by CORINE (COoRdination of INformation on the Environment) project \citep{EEA}, which integrates remote sensing images and photo interpretation to produce a dataset classifying the spatial units (pixels) into 44 land use classes. The coordinate system is EPSG:4326 from the World Geodetic System 1984, used in GPS. The datasets are made of pixels of size 250$\times$250 metres. Guidelines are then provided to dichotomize the dataset into urban and non-urban pixels, transforming land use data in Urban Morphological Zone (UMZ) data. An Urban Morphological Zone can be defined as `a set of urban areas laying less than 200m apart' \citep{EEA}. The Corine Land Cover classes used to build the Urban Morphological Zone dataset are: `Continuous urban fabric', `Discontinuous urban fabric', `Industrial or commercial units', `Green urban areas'. Moreover, `Port areas', `Airports', `Road and rail networks' and `Sport and leisure facilities' are also considered if they are neighbours to the core classes. UMZ data are useful to identify shapes and patterns of urban areas, and thus to detect urban sprawl \citep{nostro-ecolin}. Data are available for years 1990, 2006 and 2012; we selected the first and last time point for three cities in different areas of Europe. Cities are chosen based on results in \citet{eea_foen}: this report measures sprawl at country level based on three indices which take different aspects into account. We focused on the \textit{DIS}, dispersion of built-up areas, which characterises the settlement pattern according to a geometric perspective. The first city is Eindhoven, The Netherlands, chosen because the country is classified among the highly sprawled ones. The second city is Lublin, Poland, one of the countries below the average European sprawl level. The third one is Bologna, Italy, a country with an average level of sprawl. They were selected together with their commuting belts, i.e. an extension of the urban centre when this stretches beyond the administrative city boundaries; the belts include the municipalities surrounding (i.e. sharing borders) with the main city. For Eindhoven, they are Best, Eersel, Geldrop, Heeze-Leende, Nuenen, Oirschot, Son en Breugel, Veldhoven  and Waalre. For Lublin, they are G\l usk, Jastków, Konopnica, Niedrzwica Du\.{z}a, Niemce, \'{S}widnik and Wólka. For Bologna, they are Anzola dell'Emilia, Calderara di Reno, Casalecchio di Reno, Castel Maggiore, Castenaso, Granarolo dell'Emilia, Pianoro, San Lazzaro di Savena, Sasso Marconi, Zola Predosa.
A total of six binary raster datasets is thus considered: 3 cities at 2 time points, see Figure \ref{fig:bolo}.
\begin{center}
Insert Figure \ref{fig:bolo} about here
\end{center}

Polygonal maps with administrative boundaries are superimposed over Europe for selecting the areas of interest. The three cities have a similar population and spatial extension. Indeed, the enclosing rectangle around Eindhoven is 121$\times$127 pixels, and the urbanized ones are 18\% of the total in 1990 and 25\% in 2012. The rectangle around Lublin is 167$\times$140 pixels, with 9\% urban pixels in 1990 and 16\% in 2012. Bologna's rectangle is 135$\times$124 pixels, and its percentage of urban pixels is 16\% in 1990 and 18\% in 2012.

\subsection{Batty's and \citeauthor{karlstrom}'s entropy}

The area of each city with its commuting belt is partitioned following two different criteria. The first one corresponds to the administrative boundaries of the municipalities. The second option is the analogous of the equivalent option introduced for the simulation study in Section \ref{sec:simres_batty}: it considers concentric sub-areas defined by $5$ annuli with the same width, covering the whole area and centered in the centroid of each main city.

Under the administrative boundary partition, three neighbourhood distances for \citeauthor{karlstrom}'s entropy are chosen following the same idea of the simulation study: the $5$th percentile, first quartile and median of the distribution of distances among sub-areas. For the concentric area partition, the three distances are set to include from 1 to 3 neighbouring sub-areas. This way, a total of six neighbourhood systems are considered for each city with its commuting belt.

In order to compare results, entropies are divided by their maxima indicated in Sections \ref{sec:batty} and \ref{sec:kc}.
\begin{center}
Insert Table \ref{tab:powertab} about here
\end{center}
Results in Table \ref{tab:powertab} show that, for both partitions, Batty's entropy confirms the EEA country level sprawl ranking: the area of Lublin is the less sprawled, the highest level is detected for Eindhoven and Bologna constitutes an intermediate case. Moreover, Eindhoven can be classified as a sprawled city following the reference set of Section \ref{sec:simres_batty}: its entropy values are greater than $0.985$, the relative benchmark corresponding to the decentralized configuration. When introducing neighbourhood distances for \citeauthor{karlstrom}'s entropy, this ranking is further emphasized, especially at distances $nd_1$ and $nd_2$ under both the administrative boundary and concentric area partition. Conversely, extending the neighbourhood to $nd_3$ is less informative in this case study: entropies become similar, without help in detecting urban sprawl. By comparing the results over time, urban sprawl tends to increase for all cities, especially for Lublin.

\subsection{Spatial mutual information and residual entropy}

\begin{figure}
\centering
 \caption{Proportional partial spatial mutual information (grey area) and residual entropy (white area), for the first option. From left to right, Eindhoven, Lublin and Bologna in 1990 (higher panels) and 2012 (lower panels).}    \includegraphics[width=\textwidth]{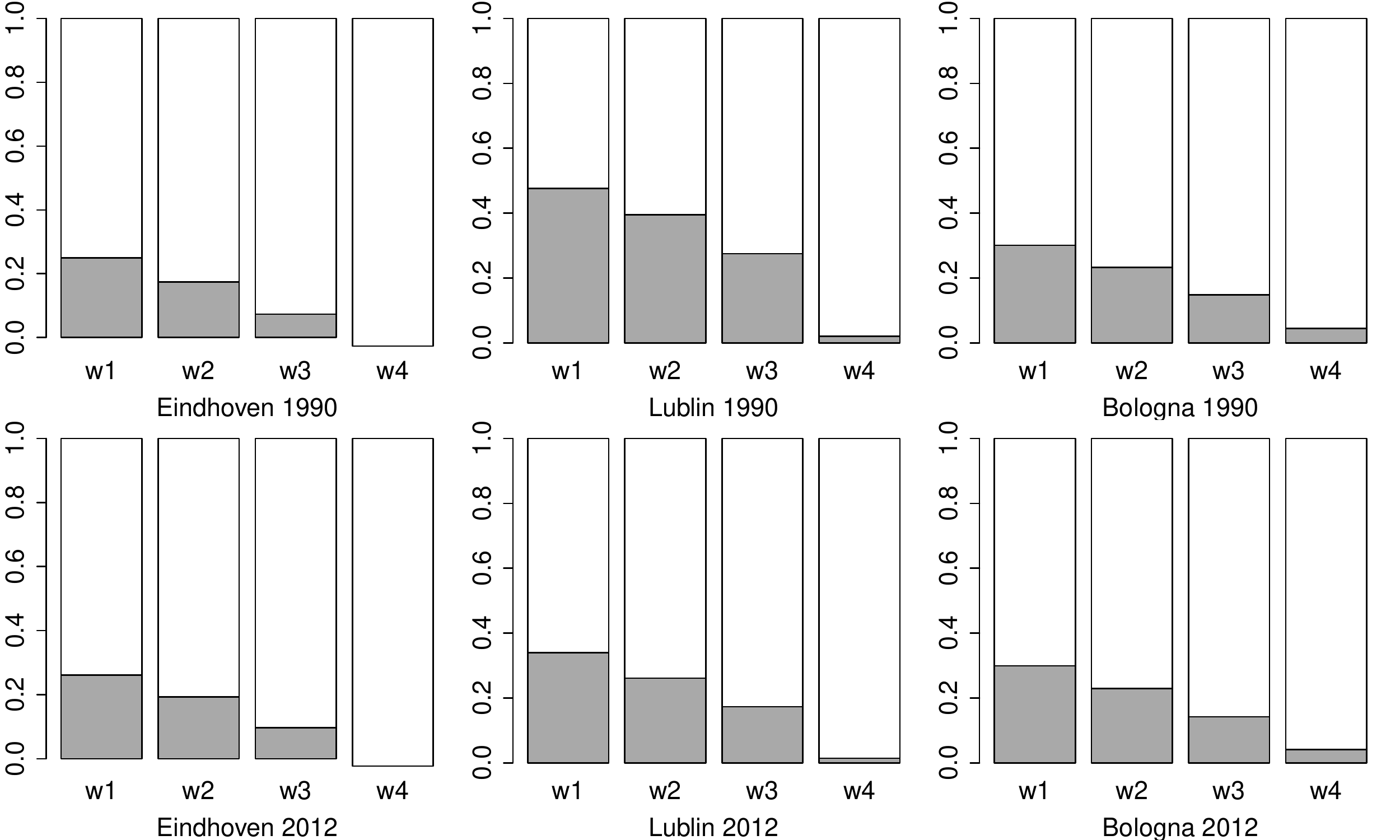}
         \label{fig:bo_spatent}
\end{figure}

Partial terms of spatial mutual information and residual entropy are computed following the same two distance options of Section \ref{sec:simres_Z}. In particular, the first one sets $w_1$ and $w_2$ to the $4$ and $12$ nearest neighbour systems, $w_3$ begins at the final point of $w_2$ and considers up to $5$ pixels along the cardinal directions, $w_4$ captures all greater distances. For the second option, the $5$th, $25$th, $50$th percentile of the empirical distribution of distances for each city is used to choose the breaks of the $4$ distance classes $w_1$ to $w_4$. All distances refers to pairs of pixels and are measured between pixel centroids.
\begin{center}
Insert Figure \ref{fig:bo_spatent} about here
\end{center}
Results are summarized in Figure \ref{fig:bo_spatent}, which plots the values of partial spatial mutual information $PI(Z|w_k)$ and partial residual entropies $H(Z|w_k)$ for the first option. To allow space and time comparisons, their proportional versions are computed by setting the sum $PI(Z|w_k) + H(Z|w_k)$ to $1$ at each distance class $w_k$. The ranking of the cities in terms of urban sprawl is more evident in $1990$ than in $2012$, again aligning with the EEA country results: Eindhoven has a low proportion of spatial information at all distances, identifying a high sprawl level; Lublin is the least sprawled, with the highest values of partial spatial information terms. Urban sprawl increases along time, and the differences in spatial mutual information and residual entropy terms across cities become almost negligible. The most informative distance classes for detecting urban sprawl are again the smallest ones. At higher distances, spatial mutual information terms decrease and the sprawl level is difficult to assess. By considering the distributions for the three scenarios identified in Section \ref{sec:simres_Z}, at distance $w_1$ the partial mutual information of Lublin belongs to the range of values of a monocentric city; with the same criterion, Eindhoven has a decentralized configuration; finally, Bologna's partial mutual information is in the lowest tail of the distribution for a polycentric city.

Results for the second distance option (not shown) are not useful to detect and compare the urban sprawl of the three cities over space and time. Indeed, the partial terms of spatial mutual information are all very low. This is due to the fact that the most informative distances have already been declared to be the smallest ones. This cannot be appreciated with the second distance option, which is not a proper choice for the problem at hand.

\section{Concluding remarks}
\label{sec:disc}

In this work, the approaches proposed by \cite{batty76}, \cite{karlstrom} and \cite{nostro} are employed to quantify the level of urban sprawl, i.e. the chaotic expansion of cities, and their properties are assessed with a comparative study. 

From the theoretical point of view, \cite{batty76} and \cite{karlstrom}'s approach represents an interesting proposal because of the LISA-type properties, however it requires a dichotomous (or dichotomized) variable, focuses on a single definition of neighbourhood and is affected by the choice of the area partition. The advantages of spatial mutual information and spatial residual entropy of \cite{nostro} lie in the possibility of managing variables with any number of categories, decomposing the entropy due to space from that due to other sources of heterogeneity, investigating the global values and the partial terms jointly, to identify the role of space for different distance ranges. 

The comparative study of Section \ref{sec:sim} and the application of Section \ref{sec:applic} highlight the ability of both approaches to distinguish among urban patterns and detect urban sprawl. In particular, Batty's entropy allows to obtain non overlapping distributions which can be used as a reference set for classifying sprawl in real studies. Spatial mutual information and residual entropy enrich results by jointly quantifying in proportional terms the level of urban sprawl at different distance ranges, and without the need of area partitions. Some conclusive points, according to both approaches, derive from the case study of Eindhoven, Lublin and Bologna. Firstly, the EEA country ranking in terms of dispersion of built up areas is reproduced here at a city level: Lublin is the least sprawled, Bologna has an intermediate level of sprawl and Eindhoven is the most sprawled. Secondly, the situation of Eindhoven is the most critical, since its entropy values belong to the range of values of the decentralized pattern. Thirdly, all cities become more affected by the sprawl issue over time, denoting a negative urban expansion from 1990 to 2012. The selected spatial entropy measures allow both an absolute classification of cities in terms of urban sprawl, and comparison across space and time via their relative versions. This is a desirable feature of such measures, which represent a contribution to the diffusion of intuitive, easily interpretable and comparable results regarding the phenomenon of urban sprawl. 

In the study of urban sprawl, the most interesting distances are the smallest ones. At this regard, spatial mutual information and spatial residual entropy are very flexible, as they can focus on the most informative distance range to interpret the phenomenon under study. The distance classes must be suitably proposed according to the context, as shown by the different options checked in Section \ref{sec:sim} and \ref{sec:applic}. The focus on small distances is not an issue for the set of spatial entropy measure, as the choice of the classes does not affect the global result; the theoretical framework illustrated in this paper shows that, when distance classes change, these measures can be easily, rapidly and intuitively adapted.

When working with data, one should use the finest available resolution, i.e. points if data are a point pattern, or the finest grid provided if data are lattice; this is the case in the present paper. Pixel aggregation is not recommended unless motivated, as it may reduce precision in the results and requires expertise in classifying the new pixel according to land use classes. 

These well performing measures capture the spatial aspect of the complex phenomenon of dispersed urbanization; they can be integrated with other indices in order to obtain a comprehensive quantification of sprawl. This helps in focusing on the worst developed areas and contributes to solving environmental issues such as dangers to ecosystems, forest destruction, pollution and climate change.

\bigskip\noindent\textbf{Acknowledgements}\\
This work is developed under the PRIN2015 supported project 'Environmental processes and human activities: capturing their interactions via statistical methods (EPHASTAT)' [grant number 20154X8K23] funded by MIUR (Italian
Ministry of Education, University and Scientific Research).

\bibliographystyle{chicago}
\bibliography{bibdatabase_entropy}

\newpage
\begin{table}[ht]
\caption{Results for Batty's and \citeauthor{karlstrom}'s (KC) entropy with two partition options}
\footnotesize
\centering
\begin{tabular}{|c|cc|cc|cc|cc|}
\cline{1-9}
\multicolumn{9}{|c|}{Administrative boundary partition}\\
\cline{1-9}
& \multicolumn{2}{c|}{Batty} & \multicolumn{2}{c|}{KC - $nd_1$} & \multicolumn{2}{c|}{KC - $nd_2$} & \multicolumn{2}{c|}{KC - $nd_3$} \\
\cline{2-9}
& 1990 & 2012 & 1990 & 2012 & 1990 & 2012 & 1990 & 2012  \\
\cline{1-9}
Eindhoven & 0.987 & 0.990 & 0.784 & 0.814 & 0.931 & 0.933 & 0.946 & 0.948 \\
Lublin & 0.955 & 0.978 & 0.530 & 0.776 & 0.865 & 0.989 & 0.993 &  0.990  \\
Bologna & 0.980 & 0.983 & 0.766 & 0.804 & 0.869 & 0.881 & 0.935 & 0.937  \\
\cline{1-9}
\multicolumn{9}{|c|}{Concentric area partition}\\
\cline{1-9}
& \multicolumn{2}{c|}{Batty} & \multicolumn{2}{c|}{KC - $nd_1$} & \multicolumn{2}{c|}{KC - $nd_2$} & \multicolumn{2}{c|}{KC - $nd_3$} \\
\cline{2-9}
& 1990 & 2012 & 1990 & 2012 & 1990 & 2012 & 1990 & 2012 \\
\cline{1-9}
Eindhoven & 0.988 & 0.991 & 0.660 & 0.670 & 0.777 & 0.787 & 0.861 & 0.861  \\
Lublin & 0.953 & 0.982 & 0.536 & 0.623 & 0.683 & 0.772 & 0.683 & 0.861  \\
Bologna & 0.976 & 0.982 & 0.594 & 0.640 & 0.683 & 0.778 & 0.683 & 0.861  \\
\cline{1-9}
\end{tabular}
\label{tab:powertab}
\normalsize
\end{table}

\newpage

\end{document}